\documentclass[12pt]{article}

\usepackage{amsmath,amsfonts,amssymb,latexsym}

\numberwithin{equation}{section}
\def\be{\begin{equation}}
\def\ee{\end{equation}}
\def\bq{\begin{eqnarray}}
\def\eq{\end{eqnarray}}
\def\beq{\begin{eqnarray*}}
\def\eeq{\end{eqnarray*}}

\def\a{\alpha}
\def\b{\beta}

\begin{document}
\begin{titlepage}
\begin{flushright}
CERN-PH-TH/2010-083
\end{flushright}

\vspace{0.3cm}

\begin{center}
{\Huge Brane singularities with a scalar field bulk}

\vspace{1cm}

{\large Ignatios Antoniadis$^{1,*,3}$, Spiros Cotsakis$^{2,\dagger}$, Ifigeneia Klaoudatou$^{2,\ddagger}$}\\

\vspace{0.5cm}

$^1$ {\normalsize {\em Department of Physics, CERN - Theory Division}}\\
{\normalsize {\em CH--1211 Geneva 23, Switzerland}}\\

\vspace{2mm}

$^2$ {\normalsize {\em Research Group of Cosmology, Geometry and
Relativity}}\\ {\normalsize {\em Department of Information and
Communication Systems Engineering}}\\ {\normalsize {\em University
of the Aegean}}\\ {\normalsize {\em Karlovassi 83 200, Samos,
Greece}}\\
\vspace{2mm}
{\normalsize {\em E-mails:} $^*$\texttt{ignatios.antoniadis@cern.ch}, $^\dagger$\texttt{skot@aegean.gr},
$^\ddagger$\texttt{iklaoud@aegean.gr}}
\end{center}

\vspace{0.7cm}

\begin{abstract}
\noindent The singularity structure and the corresponding asymptotic
behavior of a 3-brane  coupled to a scalar field in a five
dimensional bulk is analyzed in full generality, using the method of
asymptotic splittings. It is shown that the collapse singularity at
a finite distance from the brane can be avoided only at the expense of making 
the brane world-volume 
positively or negatively curved.

\end{abstract}

\vspace{1cm}
\begin{center}
{\line(5,0){280}}
\end{center}
$^3${\small On leave from {\em CPHT (UMR CNRS 7644) Ecole Polytechnique, F-91128 Palaiseau}}

\end{titlepage}

\section{Introduction}

Some time ago, an interesting idea to address the cosmological constant problem was proposed, based on the so-called self-tuning mechanism~\cite{nima,silver}. The simplest model consists of a 3-brane embedded in a five-dimensional bulk, in the presence of a scalar field. The later is coupled to the brane in a particular way, motivated by string theory, that allows flat brane world-volume solution independently of the brane tension value. It was however realized that a singularity appears in the bulk, at some finite distance from the brane, which can also be thought as a reservoir through which the vacuum energy decays.

An obvious question is then whether the development of such a singularity is a generic feature of these models, or under what conditions may be avoided. Here and in two subsequent papers, we investigate this question in a generalized class of models. Since in this case a general solution cannot be found analytically, we use a powerful tool developed a few years ago, called method of asymptotic splittings, that allows to compute all possible asymptotic behaviors of the equations of motion around the assumed location of a singularity~\cite{skot}. Our main result is twofold:
\begin{itemize}
\item The existence of a singularity at a finite distance is unavoidable in all solutions with a flat brane. This confirms and extends the results of earlier works that made similar investigations in different models, using other methods~\cite{Gubser,Forste}.
\item The singularity can be avoided ({\em e.g.} moved at infinite distance) in several cases where the brane becomes curved, either positively or negatively. Thus, requiring absence of singularity brings back the cosmological constant problem, since the brane curvature depends on its tension that receives quartically divergent quantum corrections.
\end{itemize}
As mentioned already, our results are established in an extended version of the simplest model, where the bulk scalar field has a general coupling to the brane, motivated for instance from a loop corrected string effective action, allowing for curved world-volume. Besides the scalar field (or in the absence of it), we consider a generic bulk matter content parametrized by a fluid with an arbitrary equation of state.
For pedagogical purposes, our analysis is separated in three parts contained in three different publications. A preliminary version of our results was published in~\cite{iklaoud_6}.


In this first paper, we give a detailed picture of the dynamical evolution of
an extended version of the simplest model to include curved branes. We show that the
emergence of the finite-distance singularity is the \emph{only} possible asymptotic behavior
for a flat brane, whereas for a curved brane the singularity is
shifted at an infinite distance. We provide a detailed study of the
asymptotics of this model using the method of asymptotic splittings
expounded in \cite{skot}.

The structure of this paper is as follows: In Section 2, we
derive the form of the dynamical system on which our subsequent
asymptotic analysis is applied. In Sections 3 and 4, we give the
asymptotics of the models consisting of flat and curved brane
respectively. In Section 5 we conclude and also comment on possible
future work in various directions, considering for instance other
forms of matter in the bulk. In the Appendix, we briefly outline the
basic steps of the method of asymptotic splittings.
\section{Dynamics of scalar field-brane configuration}
In this Section we set up the basic equations for our braneworld. We
study a braneworld model consisting of a three-brane embedded in a
five-dimensional bulk space with a scalar field minimally coupled to
the bulk.  The total action $S_{total}$ splits in two parts, namely,
the bulk action $S_{bulk}$ and the brane action $S_{brane}$, \be
\label{s_tot} S_{total}=S_{bulk}+S_{brane}, \ee with \bq
S_{bulk}&=&\int d^{4}x
dY\sqrt{\textrm{det}g_{5}}\left(\frac{R}{2\kappa^{2}_{5}}-
\frac{\lambda}{2}(\nabla\phi)^{2}\right),\\
S_{brane}&=&-\int d^{4}x\sqrt{\textrm{det}g_{4}}f(\phi),\,\,
\textrm{at}\,\,Y=Y_{\ast}, \eq where $Y$ denotes the fifth bulk
dimension, $Y_{\ast}$ is the assumed initial position of the brane,
$\lambda$ is a parameter defining the type of scalar field $\phi$,
$\kappa^{2}_{5}=M_{5}^{-3}$, $M_{5}$ being the five-dimensional
Planck mass, and $f(\phi)$ denotes the tension of the brane as a
function of the scalar field.

Varying the total action (\ref{s_tot}) with respect to $g^{AB}$, we
find the five-dimensional Einstein field equations in the form
\cite{iklaoud_6}, \be \label{einst5d}
R_{AB}-\frac{1}{2}g_{AB}R=\lambda \kappa_{5}^{2} \left(\nabla
_{A}\phi\nabla_{B}\phi-\frac{1}{2}g_{AB}(\nabla\phi)^{2}\right)
+\frac{2\kappa_{5}^{2}}{\sqrt{\textrm{det}g_{5}}} \frac{\delta
(\sqrt{\textrm{det}g_{4}}f(\phi))} {\delta
g^{\a\b}}\delta_{A}^{\a}\delta_{B}^{\b}\delta(Y), \ee while the
scalar field equation is obtained by variation of the action
(\ref{s_tot}) with respect to $\phi$ \cite{iklaoud_6} and it is: \be
\label{scalarbr} \lambda \Box
_{5}\phi=-\frac{1}{\sqrt{\textrm{det}g_{5}}}\frac{\delta
(\sqrt{\textrm{det}g_{4}}f(\phi))} {\delta\phi}\delta(Y), \ee where
$A,B=1,2,3,4,5$ and $\a,\b=1,2,3,4$ while $\delta (Y)=1$ at
$Y=Y_{\ast}$ and vanishing everywhere else, and \be \Box _{5}\phi=
\frac{1}{\sqrt{\textrm{det}g_{5}}}\nabla_{A}(\sqrt{\textrm{det}g_{5}}
g^{AB}\nabla_{B}\phi) .
\ee

In the following we assume a bulk metric of the form \be
\label{warpmetric} g_{5}=a^{2}(Y)g_{4}+dY^{2}, \ee where $g_{4}$ is
the four dimensional flat, de Sitter or anti de Sitter metric, i.e.,
\be \label{branemetrics} g_{4}=-dt^{2}+f^{2}_{k}g_{3}, \ee where \be
g_{3}=dr^{2}+h^{2}_{k}g_{2} \ee and \be
g_{2}=d\theta^{2}+\sin^{2}\theta d\varphi^{2}. \ee Here $
f_{k}=1,\cosh (H t)/H,\cos (H t)/H $ ($H^{-1}$ is the de Sitter
curvature radius) and $ h_{k}=r,\sin r,\sinh r $, respectively.

The field equations (\ref{einst5d})-(\ref{scalarbr}) then take the form
\bq
\label{feq1}
\frac{a'^{2}}{a^{2}}&=&\frac{\lambda\kappa^{2}_{5}\phi'^{2}}{12}+\frac{k
H^{2}}{a^{2}}, \\
\label{feq2}
\frac{a''}{a}&=&-\frac{\lambda\kappa^{2}_{5}\phi'^{2}}{4}, \\
\label{feq3} \phi''+4\frac{a'}{a}\phi'&=&0, \eq where a prime
denotes differentiation with respect to $Y$ and $k=0,\pm 1$. The
variables to be determined are $a$, $a'$ and $\phi'$. These three
equations are not independent since Eq. (\ref{feq2}) was derived
after substitution of Eq. (\ref{feq1}) in the field equation
$G_{\a\a}=\kappa_{5}^{2}T_{\a\a}$, $\a=1,2,3,4$, \be
\frac{a''}{a}+\frac{a'^{2}}{a^{2}}-\frac{kH^{2}}{a^{2}}
=-\lambda\kappa_{5}^{2}\frac{\phi'^{2}}{6}. \ee
In our analysis below we use the independent equations (\ref{feq2}) and
(\ref{feq3}) to determine the unknown variables $a$, $a'$ and
$\phi'$, while Eq. (\ref{feq1}) will then play the role of a
constraint equation for our system.

Assuming a $Y\rightarrow -Y$ symmetry and solving the Eqs.
(\ref{einst5d}) (the -$\a\a$- component, $\a=1,2,3,4$) and
(\ref{scalarbr}) on the brane we get \bq \label{bound1}
a'(Y_{\ast})&=&-\frac{\kappa_{5}^{2}}{6}f(\phi(Y_{\ast}))a(Y_{\ast}),\\
\label{bound2}
\phi'(Y_{\ast})&=&\frac{f'(\phi(Y_{\ast}))}{2\lambda}. \eq The
particular coupling used in \cite{nima} allows only for flat
solutions to exist.
This easily follows by using equations (\ref{bound1}) and
(\ref{bound2}) and solving the FRW equation (\ref{feq1}) on the
brane for $kH^{2}$:
$$kH^{2}=\frac{a^{2}(Y_{\ast})\kappa^{2}_{5}}{12}
\left(\frac{\kappa_{5}^{2}}{3}f^{2}(\phi(Y_{\ast}))-
\frac{f'^{2}(\phi(Y_{\ast}))}{4\lambda}\right).$$ Clearly, $k$ is
identically zero if and only if:
$$\frac{f'(\phi)}{f(\phi)}=2\sqrt{\frac{\lambda}{3}}\kappa_{5},$$ or equivalently, if and
only if $f(\phi)\propto e^{2\sqrt{\lambda/3}\kappa_{5}\phi}$ (the
authors of \cite{nima} have set $\lambda=3$ and hence the
appropriate choice for the brane tension in that case is
$f(\phi)\propto e^{2\kappa_{5}\phi}$). In our more general problem,
the coupling function cannot be fixed this way. By working with
other couplings we can allow for non-flat, maximally symmetric
solutions to exist and avoid having the singularity at a finite
distance away from the position of the brane.

For the rest of this paper our purpose is to find all possible
asymptotic behaviours around the assumed position of a singularity,
denoted by $Y_{s}$, emerging from general or particular solutions of
the system (\ref{feq1})-(\ref{feq3}). The most useful tool for this
analysis is the method of asymptotic splittings \cite{skot} (see the
Appendix for a brief introduction) in which we start by setting \be
x=a, \quad y=a', \quad z=\phi', \ee The field equations (\ref{feq2})
and (\ref{feq3}) become the following system of ordinary
differential equations: \bq \label{syst1_1}
x'&=&y \\
\label{syst1_2}
y'&=&-\lambda Az^{2}x \\
\label{syst1_3} z'&=&-4y\frac{z}{x}, \eq where $A=\kappa^{2}_{5}/4$.
Hence, we have a dynamical system determined by the non-polynomial
vector field \be \mathbf{f}=\left(y,-\lambda
Az^{2}x,-4y\frac{z}{x}\right)^{\intercal}. \ee Equation (\ref{feq1})
does not include any terms containing derivatives with respect to
$\Upsilon$; it is a constraint equation which in terms of the new
variables takes the form \be \label{constraint1}
\frac{y^{2}}{x^{2}}=\frac{A\lambda}{3} z^{2}+\frac{k H^{2}}{x^{2}}.
\ee Equations (\ref{syst1_1})-(\ref{syst1_3}) and
(\ref{constraint1}) constitute the basic dynamical system of our
study. There are two major cases to be treated, the first is when we
choose $k=0$ in (\ref{constraint1}) and corresponds to a brane being
flat, while in the second case $k\neq 0$, giving constant curvature
to the brane. We shall treat these two cases independently in what
follows. One important result of our analysis of this system will be
that the inclusion of nonzero curvature for the brane moves the
singularity an infinite distance away from the brane.
\section{Flat brane: Finite-distance singularity}
In this Section we take $k=0$ in the basic constraint equation \be
\label{constraint_flat} \frac{y^{2}}{x^{2}}=\frac{A\lambda}{3}
z^{2}. \ee We shall show that the only possible asymptotic behaviour
of the solutions of this system (flat brane) is that $a\rightarrow
0$, $a'\rightarrow \infty$ and $\phi'\rightarrow \infty$, as
$Y\rightarrow Y_{s}$.

We start our asymptotic analysis by inserting the forms \be
\label{dominant forms} (x,y,z)=(\alpha\Upsilon^{p},\beta
\Upsilon^{q},\delta\Upsilon^{r}), \ee in the system
(\ref{syst1_1})-(\ref{syst1_3}), where \be (p,q,r)\in\mathbb{Q}^{3}
\quad \textrm{and} \quad (\alpha,\beta,\delta)\in
\mathbb{C}^{3}\smallsetminus\{\mathbf{0}\}. \ee We find that the
only possible dominant balance in the neighborhood of the
singularity (that is pairs of the form
$\mathcal{B}=\{\mathbf{a},\mathbf{p}\}$, where
$\mathbf{a}=(\alpha,\beta,\delta)$, $\mathbf{p}=(p,q,r)$,
determining the dominant asymptotics as we approach the singularity)
is the following balance \be \label{sing}
\mathcal{B}_{1}=\{(\alpha,\alpha/4,\sqrt{3}/(4\sqrt{A\lambda})),
(1/4,-3/4,-1)\}. \ee (A second balance $\mathcal{B}_{2}$ becomes
only possible when we allow for non-zero curvature, $k\neq 0$, and
will be analysed in the next Section. There are no other acceptable
balances, hence all the possible asymptotic behaviours for a flat
and curved brane can be described uniquely by the balances
$\mathcal{B}_{1}$ and $\mathcal{B}_{2}$ respectively.)

Let us now focus on building a series expansion in the neighborhood
of the singularity to justify the asymptotics found above. We start
by calculating the Kowalevskaya exponents, eigenvalues of the matrix
$\mathcal{K}=D\mathbf{f}(\mathbf{a})-\textrm{diag}(\mathbf{p}),$
where $D\mathbf{f}(\mathbf{a})$ is the Jacobian matrix of
$\mathbf{f}$, which in our case reads: \be D\mathbf{f}(x,y,z)=\left(
                     \begin{array}{ccc}
                       0 & 1             & 0 \\
               -A\lambda z^{2} & 0             & -2A\lambda x z \\
       \dfrac{4y z}{x^{2}} & -\dfrac{4z}{x} & -\dfrac{4y}{x} \\
                     \end{array}
                   \right),
\ee
to be evaluated on $\mathbf{a}$. For the $\mathcal{B}_{1}$ balance we have that
$\mathbf{a}=(\alpha,\alpha/4,\sqrt{3}/(4\sqrt{A\lambda}))$, and
$\mathbf{p}=(1/4,-3/4,-1)$, thus \be \mathcal{K}=\left(
                     \begin{array}{ccc}
                       -\dfrac{1}{4} & 1                                   & 0 \\
                      -\dfrac{3}{16} & \dfrac{3}{4}             & -\dfrac{\sqrt{3A\lambda}\a}{2} \\
\dfrac{\sqrt{3}}{4\a\sqrt{A\lambda}} & -\dfrac{\sqrt{3}}{\a\sqrt{A\lambda}} & 0 \\
                     \end{array}
                   \right).
\ee The $\mathcal{K}$-exponents are then given by \bq
\textrm{spec}(\mathcal{K})=\{-1,0,3/2\}. \eq These exponents
correspond to the indices of the series coefficients where arbitrary
constants first appear. The $-1$ exponent signals the arbitrary
position of the singularity, $Y_{s}$. We see that the first balance
$\mathcal{B}_{1}$ has two non-negative rational eigenvalues which
means that it describes the asymptotics of a general solution in the
form of a series expansion, i.e., a series form of the solution
having the full number of arbitrary constants (which for our system
equals to two). In order to construct an asymptotic expansion of
this solution valid in the neighborhood of the singularity, we
substitute in the system (\ref{syst1_1})-(\ref{syst1_3}) the series
expansions
$$\mathbf{x}=\Upsilon^{\mathbf{p}}(\mathbf{a}+
\Sigma_{j=1}^{\infty}\mathbf{c}_{j}\Upsilon^{j/s}),$$ where
$\mathbf{x}=(x,y,z)$, $\mathbf{c}_{j}=(c_{j1},c_{j2},c_{j3})$, and
$s$ is the least common multiple of the denominators of the positive
eigenvalues; here $s=2$, and the corresponding series expansions are
given by the following forms:
\be
x=\Sigma_{j=0}^{\infty}c_{j1}\Upsilon^{j/2+1/4},\quad
y=\Sigma_{j=0}^{\infty}c_{j2}\Upsilon^{j/2-3/4},\quad
z=\Sigma_{j=0}^{\infty}c_{j3}\Upsilon^{j/2-1}.
\ee
Therefore we arrive at the following asymptotic solution around the singularity:
\bq
\label{Puis_1x}
x&=&\alpha\Upsilon^{1/4}+\frac{4}{7}c_{32}\Upsilon^{7/4}+\cdots \\
y&=&\frac{\alpha}{4}\Upsilon^{-3/4}+c_{32}\Upsilon^{3/4}+\cdots\\
\label{Puis_1z} z&=&\frac{\sqrt{3}}{4\sqrt{A\lambda}}\Upsilon^{-1}-
\frac{4\sqrt{3}}{7\alpha\sqrt{A\lambda}}c_{32}\Upsilon^{1/2}+\cdots.
\eq

The last step is to check whether for each $j$ satisfying $j/2=\rho$ with
$\rho$ a positive eigenvalue, the corresponding eigenvector $v$ of
the $\mathcal{K}$ matrix is such that the compatibility
conditions hold, namely, \be v^{\top}\cdot P_{j}=0, \ee where
$P_{j}$ are polynomials in $\mathbf{c}_{i},\ldots, \mathbf{c}_{j-1}$
given by
\be
\mathcal{K}\mathbf{c}_{j}-(j/s)\mathbf{c}_{j}=P_{j}.
\ee
Here the relation $j/2=3/2$ is valid only for
$j=3$ and the associated eigenvector is \be
\upsilon=\left(-\frac{\a\sqrt{A\lambda}}{\sqrt{3}},
-\frac{7\a\sqrt{A\lambda}}{4\sqrt{3}},1\right). \ee The
compatibility condition,
\be
\upsilon\cdot
(\mathcal{K}-(3/2)\mathcal{I}_{3})\mathbf{c}_{3}=0,
\ee
therefore indeed holds since
\be
(\mathcal{K}-(3/2)\mathcal{I}_{3})\mathbf{c}_{3}=c_{32} \left(
  \begin{array}{ccc}
    -\dfrac{7}{4} & 1 & 0 \\ \\
    -\dfrac{3}{16} & -\dfrac{3}{4} & -\dfrac{\a\sqrt{3A\lambda}}{2} \\ \\
    \dfrac{\sqrt{3}}{4\a\sqrt{A\lambda}} & -\dfrac{\sqrt{3}}{\a\sqrt{A\lambda}} & -\dfrac{3}{2} \\
  \end{array}
\right) \left(
  \begin{array}{c}
    \dfrac{4}{7} \\ \\
    1 \\ \\
     -\dfrac{4\sqrt{3}}{7\a\sqrt{A\lambda}}\\
  \end{array}
\right)=\left(
          \begin{array}{c}
            0 \\ \\
            0 \\ \\
            0 \\
          \end{array}
        \right).
\ee
This shows that a representation of the solution asymptotically
by a Puiseux series as given in Eqs. (\ref{Puis_1x})-(\ref{Puis_1z}) is valid.
Hence we conclude that near
the singularity at finite distance $Y_{s}$ from the brane the
asymptotic forms of the variables are:
\be \label{behscI}
a\rightarrow 0, \quad a'\rightarrow\infty, \quad \phi'\rightarrow
\infty.
\ee
This is exactly the asymptotic behaviour of the solution
found previously by Arkani-Hammed \emph{et al} in \cite{nima}.
Our analysis shows that this is \emph{the only possible} asymptotic behaviour
for a flat brane since there exist no other dominant balances in this case.
\section{Curved brane: Infinite-distance singularity}
In this Section we show that the collapse singularity that necessarily arises
in the case of a flat brane is avoided (or shifted at an infinite distance
away from the brane) when we consider a curved brane instead.

The new asymptotics follow from the study of a second balance that results from
the substitution of (\ref{dominant forms}) in (\ref{syst1_1})-(\ref{syst1_3}).
We calculate this new balance to be,
\be
\label{nonsing}
\mathcal{B}_{2}=\{(\alpha,\alpha,0),(1,0,-1)\}.
\ee
It corresponds to a particular solution for a \emph{curved brane}
since it satisfies Eq. (\ref{constraint1}) for $k\neq 0$ and
$\alpha^{2}=k H^{2}$ (here we have to sacrifice one arbitrary constant
by setting it equal to $kH^{2}$), $k=\pm 1$. The $\mathcal{K}$-matrix of
$\mathcal{B}_{2}$ is
\be
\mathcal{K}=D\mathbf{f}((\a,\a,0))-\textrm{diag}(1,0,-1)=\left(
  \begin{array}{ccc}
    -1 & 1 &  0 \\
    0 & 0 &  0 \\
    0 & 0 & -3 \\
  \end{array}
\right),
\ee
with eigenvalues
\be
\textrm{spec}(\mathcal{K})=\{-1,0,-3\}.
\ee
Thus for the balance $\mathcal{B}_{2}$ we find two distinct negative
integer $\mathcal{K}$-exponents and an infinite
expansion in negative powers of a \emph{particular} solution (recall that we
had to sacrifice one arbitrary constant) around the presumed
singularity at $Y_{s}$,
with the negative $\mathcal{K}$-exponents signaling the positions
where the arbitrary constants first appear \cite{fordy}. We
therefore expand the variables in series with descending powers of
$\Upsilon$ in order to meet the two arbitrary constants occurring
for $j=-1$ and $j=-3$, i.e., \be
x=\Sigma_{j=0}^{-\infty}c_{j1}\Upsilon^{j+1}, \quad
y=\Sigma_{j=0}^{-\infty}c_{j2}\Upsilon^{j}, \quad
z=\Sigma_{j=0}^{-\infty}c_{j3}\Upsilon^{j-1}. \ee Substituting these
series expansions back in the system (\ref{syst1_1})-(\ref{syst1_3})
and after some manipulation, we find the following asymptotic
behaviour, \bq \label{Puis_2x}
x&=&\alpha\Upsilon+c_{-1\,1}+\cdots \\
y&=&\alpha+\cdots \\
\label{Puis_2z} z&=&c_{-3\,3}\Upsilon^{-4}+\cdots .
\eq
Let us check the compatibility conditions for $j=-1$ and $j=-3$.
We find that
\be
(\mathcal{K}+\mathcal{I}_{3})\mathbf{c}_{-1}=\left(
                                                \begin{array}{ccc}
                                                  0 & 1 &  0 \\
                                                  0 & 1 &  0 \\
                                                  0 & 0 & -2 \\
                                                \end{array}
                                              \right)
\left(
  \begin{array}{c}
    c_{-11} \\
    0 \\
    0 \\
  \end{array}
\right)=\left(
          \begin{array}{c}
            0 \\
            0 \\
            0 \\
          \end{array}
        \right),
\ee
and
\be
(\mathcal{K}+3\mathcal{I}_{3})\mathbf{c}_{-3}=\left(
                                                \begin{array}{ccc}
                                                  2 & 1 & 0 \\
                                                  0 & 3 & 0 \\
                                                  0 & 0 & 0 \\
                                                \end{array}
                                              \right)
\left(
  \begin{array}{c}
    0 \\
    0 \\
    c_{33} \\
  \end{array}
\right)=\left(
          \begin{array}{c}
            0 \\
            0 \\
            0 \\
          \end{array}
        \right),
\ee so that the compatibility conditions are indeed satisfied. The
expansions given by Eqs. (\ref{Puis_2x})-(\ref{Puis_2z}) are
therefore valid, and we can say that as $\Upsilon\rightarrow 0$, or
equivalently as $S\equiv 1/\Upsilon\rightarrow \infty$, we have that
\be \label{behscII} a\rightarrow \infty, \quad a'\rightarrow \infty,
\quad \phi'\rightarrow \infty. \ee Therefore for a curved brane we
find that there can be no finite-distance singularities. The only
possible asymptotic behaviour is the one given in (\ref{behscII})
which is only valid at an infinite distance from the brane.
\section{Conclusions}
In this paper we studied a braneworld consisting of a three-brane
embedded in a five-dimensional bulk space filled with a scalar field
with a special emphasis in the possible formation of finite-distance
singularities away from the brane into the bulk.
We have shown that the dynamical behaviour of this model strongly depends
on the spatial geometry of the brane, in particular whether it is flat or not.
For a flat brane the model experiences a finite-distance singularity toward
which all the vacuum energy decays (since $\phi'\rightarrow\infty$,
as $Y\rightarrow Y_{s}$), whereas for a curved brane the model avoids the
singularity which is now located at an infinite distance.

It is interesting that a third balance which initially arises from
the substitution of (\ref{dominant forms}) in
(\ref{syst1_1})-(\ref{syst1_3}), namely, the form
$$\mathcal{B}_{3}=\{(\alpha,0,0),(0,-1,-1)\},$$ is not acceptable
for the model we consider in this paper since it does not give the
necessary $-1$ $\mathcal{K}$-exponent.
In future work \cite{akc2} and \cite{akc3}, we will see that this
balance although impossible in the case treated here, \emph{does}
become possible (although in a somewhat `mild' form) when we replace
the scalar field studied here with other matter components such as a
perfect fluid or a combination of a perfect fluid and a scalar
field. We  therefore conclude that for the case of interest in this
paper the collapse singularity found is the only type of singularity
that can develop at a finite distance from a flat brane. 

\section*{Acknowledgements}
S.C. and I.K. are grateful to CERN-Theory Division, where part of
their work was done, for making their visits there possible and for
allowing them to use its excellent facilities. The work of I.A. was
supported in part by the European Commission under the  ERC Advanced
Grant 226371 and the contract PITN-GA-2009-237920 and in part by the
CNRS grant GRC APIC PICS 3747.

\appendix
\section{Appendix: The method of asymptotic splittings}
We refer briefly here to the basic steps of the method of asymptotic
splittings. A detailed analysis can be found in Ref. \cite{skot}.

Consider a system of $n$ first order ordinary differential equations
\be \label{arb_syst} \mathbf{x'}=\mathbf{f}(\mathbf{x}), \ee where
$\mathbf{x}=(x_{1},\ldots,x_{n})\in \mathbb{R}^{n}$,
$\mathbf{f}(\mathbf{x})=(f_{1}(\mathbf{x}),\ldots,f_{n}(\mathbf{x}))$
and $'\equiv\frac{d}{dY}$, $Y$ being the independent variable. In
this paper, we refrain from calling $Y$ a time variable and giving
it the interpretation of time. Since we are interested in
singularities located at a \emph{distance} from the brane and into
the bulk, it seems more appropriate to talk about finite-distance
singularities and give to the $Y$ variable a spatial interpretation.
The general solution of the above system contains $n$ arbitrary
constants and describes all possible behaviours of the system
starting from arbitrary initial data. Any particular solution of
(\ref{arb_syst}), on the other hand, contains less than $n$
arbitrary constants and describes a possible behaviour of the system
emerging from a proper subset of initial data space.

We say that a solution of the dynamical system (\ref{arb_syst})
exhibits a finite-distance singularity if there exists a $Y_{s}\in
\mathbb{R}$ and a $\mathbf{x}_{0}\in \mathbb{R}^{n}$ such that
\be \lim_{Y\rightarrow
Y_{s}}\|\mathbf{x}(Y;\mathbf{x}_{0})\|\rightarrow\infty,
\ee where $\|\centerdot\|$ is any $L^{p}$ norm. The purpose of
singularity analysis (cf. \cite{skot},
\cite{goriely}) 
is to build series expansions of solutions around the presumed
position of a singularity at $Y_{s}$ in order to study the different
asymptotic behaviours of the solutions of the system
(\ref{arb_syst}) as one approaches this singularity. In particular,
we look for series expansions of solutions that take the form of a
Puiseux series (any $\log$ terms absent), namely, a series of the form
\be \label{Puiseux}
\mathbf{x}=\Upsilon^{\mathbf{p}}\left(\mathbf{a}
+\Sigma_{i=1}^{\infty}\mathbf{c}_{i}\Upsilon^{i/s}\right), \ee where
$\Upsilon=Y-Y_{s}$, $\mathbf{p}\in \mathbb{Q}^{n}$,
$s\in\mathbb{N}$.

The method of asymptotic splittings for any system of the form
(\ref{arb_syst}) is realized by taking the following steps:

$\bullet$ First, we find all the possible \emph{weight-homogeneous
decompositions} of the vector field $\mathbf{f}$ by splitting it
into components $\mathbf{f}^{(j)}$: \be
\mathbf{f}=\mathbf{f}^{(0)}+\mathbf{f}^{(1)}+\ldots
+\mathbf{f}^{(k)}, \ee with each of these components being
\emph{weight homogeneous}, that is to say \be
\mathbf{f^{(j)}}(\mathbf{a}\Upsilon^{\mathbf{p}})=\tau^{\mathbf{p}
+\mathbf{1}(q^{(j)}-1)}\mathbf{f}^{(j)}(\mathbf{a}) \quad
j=0,\ldots,k, \ee where $\mathbf{a}\in\mathbb{R}^{n}$ and $q^{(j)}$
are the positive non-dominant exponents that are defined by
(\ref{sub_exp}) below.

$\bullet$ We substitute the forms
$\mathbf{x}=\mathbf{a}\mathbf{\Upsilon}^{\mathbf{p}}$ in the system
$\mathbf{x}'=\mathbf{f}^{(0)}(\mathbf{x})$ in order to find all
possible \emph{dominant balances}, i.e., finite sets of the form
$\{\mathbf{a},\mathbf{p}\}$. The \emph{order} of each balance is
defined as the number of the non-zero components of $\mathbf{a}$.

$\bullet$ For each of these balances we check the validity of the
following \emph{dominance condition}:
\be \label{dominance}
\lim_{\Upsilon\rightarrow 0}
\frac{\Sigma_{j=1}^{k}\mathbf{f}^{(j)}(\mathbf{a}\Upsilon^{\mathbf{p}})}
{\Upsilon^{\mathbf{p}-1}}=0, \ee and define the non-dominant
exponents $q^{(j)}$, $j=1,\ldots, k$ by the requirement that
\be
\label{sub_exp}
\frac{\Sigma_{j=1}^{k}\mathbf{f}^{(j)}(\Upsilon^{\mathbf{p}})}
{\Upsilon^{\mathbf{p}-1}}\sim\Upsilon^{q^{(j)}}. \ee The balances
that cannot satisfy the condition (\ref{dominance}) are then
discarded.

$\bullet$ We compute the Kovalevskaya matrix $\mathcal{K}$
defined by \be \mathcal{K}=D\mathbf{f}^{(0)}(\mathbf{a})-diag
\mathbf{p}, \ee where $D\mathbf{f}^{(0)}(\mathbf{a})$ is the
Jacobian matrix of $\mathbf{f}^{(0)}$ evaluated at $\mathbf{a}$.

$\bullet$ We calculate the spectrum of the $\mathcal{K}$-matrix,
$spec(\mathcal{K})$, that is the set of its $n$ eigenvalues also called
the \emph{$\mathcal{K}$-exponents}. The arbitrary constants of
any particular or general solution first appear in those terms in
the series (\ref{Puiseux}) whose coefficients $\mathbf{c}_{k}$ have
indices $k=\rho s$, where $\rho$ is a non-negative
$\mathcal{K}$-exponent and $s$ is the least common multiple of the
denominators of the set consisting of the non-dominant exponents
$q^{(j)}$ and of the positive $\mathcal{K}$-exponents (cf.
\cite{skot}, \cite{goriely}).
The number of non-negative $\mathcal{K}$-exponents equals therefore
the number of arbitrary constants that appear in the series
expansions of (\ref{Puiseux}). There is always the $-1$ exponent
that corresponds to the position of the singularity, $Y_{s}$. (A dominant
balance corresponds thus to a general solution if it possesses $n-1$ non-negative
$\mathcal{K}$-exponents (the $n$th arbitrary constant is the
position of the singularity, $Y_{s}$)).

$\bullet$ We substitute the Puiseux series: \be
x_{i}=\Sigma_{j=0}^{\infty}c_{ji}\Upsilon^{p_{i}+j/s}, i=1,\ldots,
n, \ee in the system (\ref{arb_syst}).

$\bullet$ We find the coefficients $\mathbf{c}_{j}$ by
solving the recursion relations \be
\mathcal{K}\mathbf{c}_{j}-\frac{j}{s}\mathbf{c}_{j}=
\mathbf{P}_{j}(\mathbf{c}_{1},\ldots, \mathbf{c}_{j-1}) \ee where
$\mathbf{P}_{j}$ are polynomials that are read off from the original
system.

$\bullet$ We verify that for every $j=\rho s$, with $\rho$ a
positive $\mathcal{K}$-exponent, the following compatibility
conditions hold:
\be
\label{comp_cond}
\mathbf{\upsilon}^{\intercal}\cdot\mathbf{P}_{j}=0,
\ee
where $\mathbf{\upsilon}$ is an eigenvector associated with the positive
$\mathcal{K}$-exponent $\rho$.

$\bullet$ We repeat the procedure for each possible decomposition.

We note that if the compatibility condition above (Eq. (\ref{comp_cond})) is
violated at some eigenvalue in the $\textrm{spec}(\mathcal{K})$, then the original Puiseux
series representation of the solution cannot be admitted and instead we have
to use a \emph{$\psi$-series} for each one of the eigenvalues with this property.
This is a series that includes $\log$ terms of the form
\be \mathbf{x}=\Upsilon^{\mathbf{p}}\left(\mathbf{a}
+\Sigma_{i,j=1}^{\infty}\mathbf{c}_{ij}\Upsilon^{i/s}(\Upsilon^{\rho}\log\Upsilon)^{j/s}\right),
\ee
where $\rho$ is the $\mathcal{K}$-exponent for which the
compatibility condition is violated. The rest of the procedure in
this case is the same as before.

\end{document}